\documentclass[fleqn,12pt,twoside]{article}
\usepackage[headings]{espcrc1}

\readRCS
$Id: espcrc1.tex,v 1.2 2004/02/24 11:22:11 spepping Exp $
\usepackage{graphicx}
\usepackage[figuresright]{rotating}

\newcommand{\AmS}{{\protect\the\textfont2
  A\kern-.1667em\lower.5ex\hbox{M}\kern-.125emS}}
\newcommand{\beq}{\begin{equation}}
\newcommand{\eeq}{\end{equation}}

\newcommand{\im}{\mathrm{Im}}

\newcommand{\arcsinh}{\mathrm{arcsinh}}

\def\eq#1{{(\ref{#1})}}

\title{Quantum Black Holes \\ and Thermalization in Relativistic Heavy Ion Collsions}

\author{D. Kharzeev\address[BNL]{
Nuclear Theory Group\\
Physics Department\\
Brookhaven National Laboratory\\
Upton, New York 11973, USA}
        \thanks{Work supported by the U.S. Department of Energy under Contract No. DE-AC02-98CH10886.}}
              
\runtitle{Quantum black holes and thermalization in relativistic heavy ion collisions}
\runauthor{D. Kharzeev}

\begin{document}

% typeset front matter
\maketitle

\begin{abstract}
A new thermalization scenario for heavy ion collisions is discussed. It is based on the Hawking--Unruh effect: 
an observer moving with an acceleration $a$ experiences the influence of a thermal bath with an effective 
temperature $T = a / 2\pi$, similar to the one present in the vicinity of a black hole horizon. 
In the case of heavy ion collisions, the acceleration is caused by a pulse of chromo--electric field 
$E \sim Q_s^2/g$ ($Q_s$ is the 
saturation scale, and $g$ is the strong coupling), the typical acceleration is $a \sim Q_s$, and the heat bath temperature 
is $T \simeq Q_s / 2\pi \sim 200$ MeV. In nuclear collisions at sufficiently high energies the effect can induce  a rapid thermalization over the time period of $\tau \sim \pi/Q_s $ accompanied by phase transitions. A specific 
example of chiral symmetry restoration induced by the chromo--electric field is considered; 
it is mathematically analogous to the phase transition occurring in the vicinity of a black hole. 
\end{abstract}

\section{INTRODUCTION}

There is something puzzling in the data on hadron production in various processes, from $e^+e^-$ annihilation and deep--inelastic 
scattering to heavy ion collisions: the relative abundances of different hadron species appear to follow the statistical distribution with a surprising accuracy (for reviews, see e.g. \cite{Becattini:2004td,Braun-Munzinger:2003zd}). Moreover, 
at small transverse momenta the spectra of the produced hadrons also look approximately thermal. While in heavy ion collisions it is possible 
to expect the emergence of statistical distributions as a result of intense re--interactions between the produced particles, this seems very 
implausible in $e^+ e^-$ annihilation at high energies $\sqrt{s}$, where the process of hadronization is stretched in space over a long 
distance $\sim \sqrt{s}/2 \mu^2$ and the density of produced hadrons is small ($\mu \sim \Lambda_{\rm QCD}$ is an infrared cut-off describing the hadronization scale).

Statistical distributions may emerge as a result of the saddle--point approximation to the multi--particle phase space, when the dynamics is inessential and the production 
mechanism is "phase space dominated" (see, for example,  \cite{Deinet:2005uk} and references therein). 
However, in $e^+ e^-$ annihilation this mechanism can hardly be expected to work: the jet structure, the angular distributions of the produced hadrons, and inter--jet correlations point to the all--important role of QCD dynamics of gluon radiation (for a recent review, see \cite{Dokshitzer:2004tp}), and thus the "phase space dominance" cannot be invoked. It thus looks natural to think that the emergence of statistical hadron abundances  has something to do with the process of hadronization -- in other words, with the way in which the QCD vacuum responds to the external color  fields, as was advocated by Dokshitzer  \cite{Dokshitzer:2004tp}. 

This attractive and deep idea however does not yet allow one to understand why the vacuum response to the external carriers of color field leads to the emergence of statistical distributions. To do this, we are forced to discuss the structure of the QCD vacuum.
Little is known about it, but we do know that QCD vacuum is populated by the fluctuations of color fields, some of which are of semi--classical nature \cite{Belavin:1975fg}. 
It may therefore be useful to examine what is known at present 
 about the interactions of charged quanta with background classical fields. 

\vskip0.3cm

This logic will first lead us to the discussion of quantum fluctuations in the background of the gravitational 
field of a black hole, where the quantum radiation appears to be thermal as shown by Hawking \cite{Hawking:1974sw}.  It was demonstrated by Unruh \cite{unruh} that the Hawking phenomenon should be present in any non--inertial frame; indeed, this is required by Einstein's Equivalence Principle. Then we will proceed to the process of electron--positron pair production in the background electric field, analyzed in QED by Schwinger \cite{schwinger}. We will see that the Schwinger formula for the rate of $e^+e^-$ pair production in a constant electric field allows for a simple statistical interpretation; moreover, for the case of a time--dependent field pulse, the spectra of the produced particles become thermal. 

These two examples inspire the picture in which the thermal character of hadron production emerges from 
the interactions with the vacuum chromo--electric field, the strength of which in the color flux model 
is parameterized in terms of the string tension, or the slope of Regge trajectories $b$. Indeed, we will see how the well--known formula for the Hagedorn temperature  
\beq\label{hag}
 T_{Hagedorn} = {\sqrt{6} \over 4 \pi} \ {1\over \sqrt{b}} 
\eeq 
 can be derived in this way.  
 
 What happens if we create semi--classical color fields of strength 
exceeding the strength of the vacuum fields? This can be achieved in the collisions of heavy ions at high energies, which are accompanied by a short pulse of the chromo--electric field $E \sim Q_s^2 / g$ of duration 
$\tau \sim Q_s^{-1}$; here $Q_s$ is the saturation scale $Q_s$ in the Color Glass 
Condensate (for a review at this Conference, see  \cite{Itakura:2005eu}), and $g$ is the strong coupling. 
I will argue that such a strong color field induces the creation of partons with a distribution which is 
isotropic and thermal in a co--moving local frame, with an effective temperature $T_{in} \simeq Q_s / 2 \pi$. 
For high enough energies and heavy enough nuclei, the value of $T_{in}$ exceeds the Hagedorn temperature; 
the produced thermal matter is thus in the deconfined phase. The phase transition in this case can also 
be understood in a geometrical picture, in which the acceleration $a \simeq Q_s$ in the chromo--electric field 
determines the curvature of space in the non--inertial Rindler frame. This is mathematically analogous to the phase 
transition induced by the presence of a massive black hole.   This talk is based on the paper with Kirill Tuchin \cite{Kharzeev:2005iz}, to which I refer for details and references.

\section{QUANTUM FLUCTUATIONS IN THE CLASSICAL BACKGROUND}

\subsection{Black hole evaporation}

In 1974 Hawking \cite{Hawking:1974sw} demonstrated 
that black holes evaporate by quantum pair production, and behave as if they have 
an effective temperature of 
\beq
T_H = {\kappa \over 2 \pi},
\eeq
where $\kappa = 4M$ is the acceleration of gravity at the surface of a black hole of mass $M$.  The thermal 
character of the black hole radiation stems from the presence of the event horizon, which hides 
the interior of the black hole from an outside observer.  
The rate of pair production in the gravitational background of a black hole can be evaluated 
by considering the tunneling through the event 
horizon. Parikh and Wilczek \cite{Parikh:1999mf} showed that the imaginary part of the action for this classically forbidden 
process corresponds to the exponent of the Boltzmann factor describing the thermal emission\footnote{Conservation laws also imply a non-thermal correction to the emission rate  \cite{Parikh:1999mf}, possibly causing a leakage of information from the black hole.}.

 Unruh \cite{unruh} has found that a similar effect arises in a uniformly accelerated 
frame, where an observer detects an apparent thermal radiation with the temperature 
\beq \label{unruhtemp}
T_U = { a \over 2 \pi};
\eeq
($a$ is the acceleration). The event horizon in this case emerges due to the existence of causally disconnected regions of space--time, conveniently described by using the Rindler coordinates.  

\subsection{Pair production in a constant electric field}

The effects associated with a heat bath of temperature (\ref{unruhtemp}) usually are not easy to detect because 
of the smallness of the acceleration $a$ in realistic experimental conditions. For example, for the acceleration of gravity 
on the surface of Earth $g \simeq 9.8 \ {\rm m\ s^{-2}}$ the corresponding temperature is only $T \simeq 4 \times 10^{-20}  \ 
{\rm K}$. (The wavelength of the thermal radiation in this case is about 1 parsec - so it would have to be detected on a cosmic scale!) 

Much larger accelerations can be achieved in electromagnetic fields, and Bell and Leinaas \cite{Bell:1982qr} considered the possible manifestations of the Hawking--Unruh effect  in particle accelerators. They argued that the presence of an apparent heat bath can cause beam depolarization. Indeed, if the energies of spin--up $E_{{\uparrow}}$ and spin--down $E_{{\downarrow}}$ states of a particle in the magnetic field of an accelerator differ by $\Delta E = E_{{\uparrow}} - E_{{\downarrow}}$, the Hawking--Unruh effect would lead to the 
thermal ratio of the occupation probabilities
\beq\label{depol}
{N_{{\uparrow}} \over N_{{\downarrow}}} \simeq \exp\left(-{\Delta E \over T_U}\right),
\eeq
where the Unruh temperature \eq{unruhtemp} is determined 
by the particle acceleration. According to \eq{depol}, a pure polarization state 
of a particle in the accelerator is inevitably diluted by the acceleration. 

If the energy spectrum of an accelerated observer is continuous, as is the case 
for a particle of mass $m$ with a transverse (with respect to the direction 
of acceleration) momentum $p_T$, a straightforward extension of \eq{depol} leads to a 
thermal distribution in the "transverse mass" $m_T = \sqrt{m^2 + p_T^2}$:
\beq\label{tranmass}
W_m(p_T) \sim \exp\left( - {m_T \over T_U} \right).
\eeq
\vskip0.3cm
An important example is provided by the dynamics of charged particles in external electric fields. 
Consider a particle of mass $m$ and charge $e$ in an external electric field of strength $E$. 
Under the influence of the Lorentz force, it  
moves with an acceleration $a = e E / m$, and the corresponding temperature is $T_U = a / 2 \pi$. 
The Boltzman factor $\exp( - m / T_U)$ entering the particle creation rate in this case is 
\beq\label{tranmassel}
W^E_m \sim \exp\left( - {2 \ \pi  m^2 \over e E} \right). 
\eeq    
This expression looks familiar -- in fact, \eq{tranmassel} differs from the classic Schwinger result for the rate of particle production in a constant electric field only by a factor of two in the exponent. 

Is this a coincidence? To answer this question, let us have a closer look at the corresponding action:
\beq\label{actfield}
S\,=\,\int\,(\,-\,m\,ds\,-\,e\,\varphi\,dt\,)\,, 
\eeq
where $\varphi$ is the electric potential, which for a constant electric field aligned along the $x$ axis is $\varphi=-Ex$ modulo an additive constant; the invariant interval is given by $ds^2=(1-v^2(t))\,dt^2$. Using the equations of motion, we can evaluate the 
action to find 
\begin{eqnarray}\label{act}
S(\tau)&=&\int^\tau\,dt\,(-\,m\,\sqrt{1\,-\,v(t)^2}\,+\,e\,E\,x(t))\,\nonumber\\
&=& -\,\frac{m}{a}\,\arcsinh(a\,\tau)\,+\,
\frac{e\,E}{2\,a^2}\,\left(a\,\tau\,(\sqrt{1\,+\,a^2\,\tau^2}\,- 2)\,+\,\arcsinh(a\,\tau)\right)\,+
\,\mathrm{const}\,.
\end{eqnarray} 
In classical mechanics the equations of motion completely specify 
the trajectory of a uniformly accelerating particle moving under the influence of a constant force 
$\vec F = -e\nabla \varphi $. In contrast, in quantum 
theory the particle has a finite probability to be found under the 
potential barrier $V(x) =eE x$ in the classically forbidden region. Mathematically, it 
comes about since the action 
\eq{act} being an analytic function of $\tau$ has an imaginary part in the Euclidean space
\beq\label{impart}
\im \,S(\tau)\,=\,\frac{m\,\pi}{a}\,-\,\frac{e\,E\,\pi}{2\,a^2}\,=\,\frac{\pi\,m^2}{2\,e\,E}\,.
\eeq 
The imaginary part of the action \eq{act} corresponds to the motion of a particle 
in Euclidean space along the trajectory
\beq\label{euctraj}
x(t_E)\,=\,a^{-1}\,(\,\sqrt{1\,-\,a^2\,t_E^2}\,-\,1\,)\,.
\eeq
Note that unlike in Minkowski space the Euclidean trajectory is bouncing 
between the two identical points $x_a=-a^{-1}$ at $t_{E,a}=-a^{-1}$ and 
$x_b=-a^{-1}$ at $t_{E,b}=a^{-1}$, and the turning point $x_a = 0$ at $t_{E,a}=0$.  Using \eq{act} we can find the 
Euclidean action between the points $a$ and $b$; it is given by $S_E(x(t_E))= \im S = \pi m^2/2 e E $. 

It 
is well known that a quasi-classical exponent describing the decay of a metastable 
state is given by the Euclidean action of the bouncing solution, \eq{gamma1}.  
The rate of tunneling under the potential barrier in the quasi-classical approximation is thus given by 
\beq\label{gamma1}
W_m^E\,\sim\,\exp(-2\,\im S)\,=\,\exp\left(-\frac{\pi\,  m^2}{e\,E}\right)\,.
\eeq
Equation \eq{gamma1} gives the probability to produce a particle and its antiparticle (each of mass $m$) out of the vacuum by a constant electric field $E$; note that the incorrect factor of $2$ in the exponent of \eq{tranmassel} has now disappeared due to the contribution of the field term in the action \eq{actfield}. 

The ratio of the probabilities to produce states of masses $M$ and $m$ is then
\beq\label{ratprob}
\frac{W_{M}^E}{W_{m}^E}\,=\,
\exp\left(-\frac{\pi\,  (M^2\,-\,m^2)}{e\,E}\right)\,.
\eeq 
The relation \eq{ratprob} allows a dual interpretation in terms of both Unruh and Schwinger effects (see e.g. \cite{Parentani:1996gd,Gabriel:1999yz,Narozhny:2003ux,Kharzeev:2005iz} and references therein). First, consider a detector with quantum levels $m$ and $M$ moving in a constant electric field. Each level is accelerated differently, however if the splitting is not large, $M-m\ll m$ we can introduce the average acceleration of the detector 
\beq\label{averacc}
\bar a\,=\,\frac{2\,e\,E}{M\,+\,m}\,.
\eeq  
Substituting \eq{averacc} into \eq{ratprob} we arrive at 
\beq\label{ratun}
\frac{W_{M}^E}{W_{m}}\,=\,\exp\left(\frac{2\,\pi\,(M\,-\,m)}{\bar a}\right)\,.
\eeq
This expression is reminiscent of the Boltzmann weight in 
a heat bath with an effective temperature (\ref{unruhtemp}): $T = \bar a/2 \pi$. It implies that the detector is effectively immersed in a photon heat bath at temperature $T\approx e   E / \pi m$. This is the renown Unruh effect \cite{unruh}.

It is important to remember that for a constant electric field the momentum distribution of the produced charged particles 
allows a statistical interpretation only in the transverse to the field direction. However for a short pulse of an electric field 
of duration $\tau \leq m/(eE)$ the distribution becomes thermal in all three directions (see \cite{Kharzeev:2005iz} and references 
therein) -- physically, this happens because the field has not enough time to perform work on curving the momenta of the produced particles.

Let us now discuss \eq{ratprob},\eq{gamma1} from the viewpoint of a field-theoretical derivation done by Schwinger \cite{schwinger}. 
It is clear that \eq{gamma1} cannot be expanded neither in powers of the coupling $e$, nor in powers of the field $E$, and so 
cannot be reproduced in any finite order of perturbation theory. Schwinger considered the one--loop QED action 
describing the electron--positron fluctuations in the background of the external electric field $E$. He has found that 
the series in the number of external field insertions indeed diverges (is not Borel summable). As a result the action ceases to be real, 
and develops an imaginary part, similarly to the simple semi--classical example considered above. 

The Hawking--Unruh interpretation therefore appears to capture an essentially non--perturbative dynamics. Indeed, a more 
rigorous treatment (for a review, see e.g. \cite{BD}) allows to establish that   the Bogoliubov transformations relating 
the particle creation and annihilation operators in Minkowski and Rindler spaces 
describe a rearrangement of the vacuum structure which cannot be captured by 
perturbative series. 
 
%\subsection{Pair production by a pulse of electric field}

\section{Event horizon and thermalization in high energy hadronic interactions}\label{secthoriz}

\subsection{Unruh effect, Hagedorn temperature, and parton saturation}\label{hagedorn}

We are now ready to address the case of hadronic interactions at high energies, 
which is the main subject of this talk. Consider a high--energy hadron of mass $m$ 
and momentum $P$ which interacts with an external field (e.g., another hadron) and transforms into a hadronic final state of invariant mass $M \gg m$. This transformation is accompanied 
by a change in the longitudinal momentum  
\beq \label{ql}
q_L = \sqrt{E^2 - m^2} - \sqrt{E^2 - M^2} \simeq {M^2 - m^2 \over 2P},
\eeq  
and therefore by a deceleration; we assumed that the particle $m$ is relativistic, with energy $E \simeq p$.

The probability for a transition to a state with an invariant mass $M$ is given by 
\beq \label{probtot}
P(M \leftarrow m) = 2 \pi |{\cal{T}}(M \leftarrow m)|^2 \ \rho(M),
\eeq
where ${\cal T}(M \leftarrow m)$ is the transition amplitude, and $\rho(M)$ is the density of hadronic final states. 
According to the results of the previous section, we expect that under the influence 
of deceleration $a$ which accompanies the change  of momentum (\ref{ql}), 
the probability $|{\cal T}|^2$ will be determined by the Unruh effect and will be given by
\beq \label{occfac}
 |{\cal T}(M \leftarrow m)|^2 \sim \exp(-2 \pi M/a)
\eeq
in the absence of any dynamical correlations; we assume $M \gg m$.

To evaluate the density of states $\rho(M)$, let us first 
use the dual resonance model (see e.g. \cite{dealfaro}, \cite{Satz:1973zc}), in which 
\beq \label{degfac}
\rho(M) \sim \exp\left({4 \pi \over \sqrt{6}}\  \sqrt{b} \ M\right),
\eeq  
where $b$ is the universal slope of the Regge trajectories, related to the 
string tension $\sigma$ by the relation $\sigma = 1/(2\pi b)$.  

The unitarity dictates that the sum of the probabilities (\ref{probtot}) over all finite states $M$ should be finite. 
Therefore, by converting the sum into integral over $M$ one can see that the eqs (\ref{occfac}) 
and (\ref{degfac}) impose the following bound on the value of acceleration $a$:
\beq \label{hag}
{a \over 2 \pi} \equiv T \leq {\sqrt{6} \over 4 \pi} \ {1\over \sqrt{b}} \equiv T_{Hag}.
\eeq  
The quantity on the r.h.s. of (\ref{hag}) is known as the Hagedorn temperature \cite{Hagedorn:1965st} -- the 
"limiting temperature of hadronic matter" derived traditionally from hadron thermodynamics. 
In our case it stems from the existence of a "limiting acceleration" $a_0$:
\beq \label{limac}
a_0 = \sqrt{3 \over 2} \ b^{-1/2}.
\eeq 
 
The meaning of the "limiting temperature" in hadron thermodynamics is well-known: above it, 
hadronic matter undergoes a phase transition into the deconfined phase, in which the quarks and gluons 
become the dynamical degrees of freedom. To establish the meaning of the limiting acceleration (\ref{limac}), let us 
consider a dissociation of the incident hadron into a large number $n \gg 1$ of partons. In this case the phase space density (\ref{degfac}) 
can be evaluated by the saddle point method ("statistical approximation"), with the result (see e.g. \cite{keijo})
\beq \label{rhostat}
\rho(M) \sim \exp( \beta M); 
\eeq
where $\beta^{-1}$ is determined by a typical parton momentum in the center-of-mass frame of the partonic 
configuration. When interpreted in partonic language, eq(\ref{degfac}) thus implies a constant value of mean transverse 
momentum ${\bar p}_T \sim \beta^{-1} \sim b^{-1/2}$. On the other hand, in the parton saturation picture, 
the mean transverse momentum has to be associated with the "saturation scale" $Q_s$ determined by the 
density of partons in the transverse plane within the wave function of the incident hadron (or a nucleus). 
This leads to the phase space density $\rho(M) \sim \exp(M/Q_s)$. The unitarity condition and the formulae 
(\ref{occfac}), (\ref{probtot}) thus lead us to the acceleration $a = Q_s$, which can exceed (\ref{limac}), and 
to the conclusion that the final partonic states are described by a thermal distribution with the temperature 
\beq \label{tempsat}
T = {Q_s \over 2 \pi}.
\eeq  
The same result can be obtained by considering the acceleration $ a = g E / m$ of a parton with off-shellness $m \equiv \sqrt{p^2} \simeq Q_s$ in an external color field $gE \simeq Q_s^2$. It is interesting to note that to exceed the limiting acceleration 
(\ref{limac}), and thus the limiting Hagedorn temperature (\ref{hag}) for the produced hadronic matter, one has to build up 
strong color fields, exceeding $gE_0 \sim 1/b$. This is achieved by parton saturation in the Color Glass Condensate, 
when $gE \simeq Q_s^2 > gE_0$ at sufficiently high energies and/or large mass numbers of the colliding nuclei. 
Parton saturation in the initial wave functions thus seems to be a necessary pre-requisite for the emergence of a 
thermal deconfined partonic matter in the final state. 
  
The thermal distribution is built over the time period of 
\beq
t_\mathrm{therm} \sim T^{-1} = { \pi \over Q_s}.
\eeq  
As discussed above, this apparent thermalization originates from the presence of the event horizon in an accelerating frame:  
the incident hadron decelerates in an external color field, which causes the emergence of the causal horizon. 
Quantum tunneling through this event horizon then produces a thermal final state of partons, in complete analogy 
with the thermal character of quantum radiation from black holes.  

\subsection{The space--time picture of relativistic heavy ion collisions}

\begin{figure}[htb]
\noindent
\vspace{-0.3cm}
\begin{minipage}[b]{.46\linewidth}
\includegraphics[width=8.3cm]{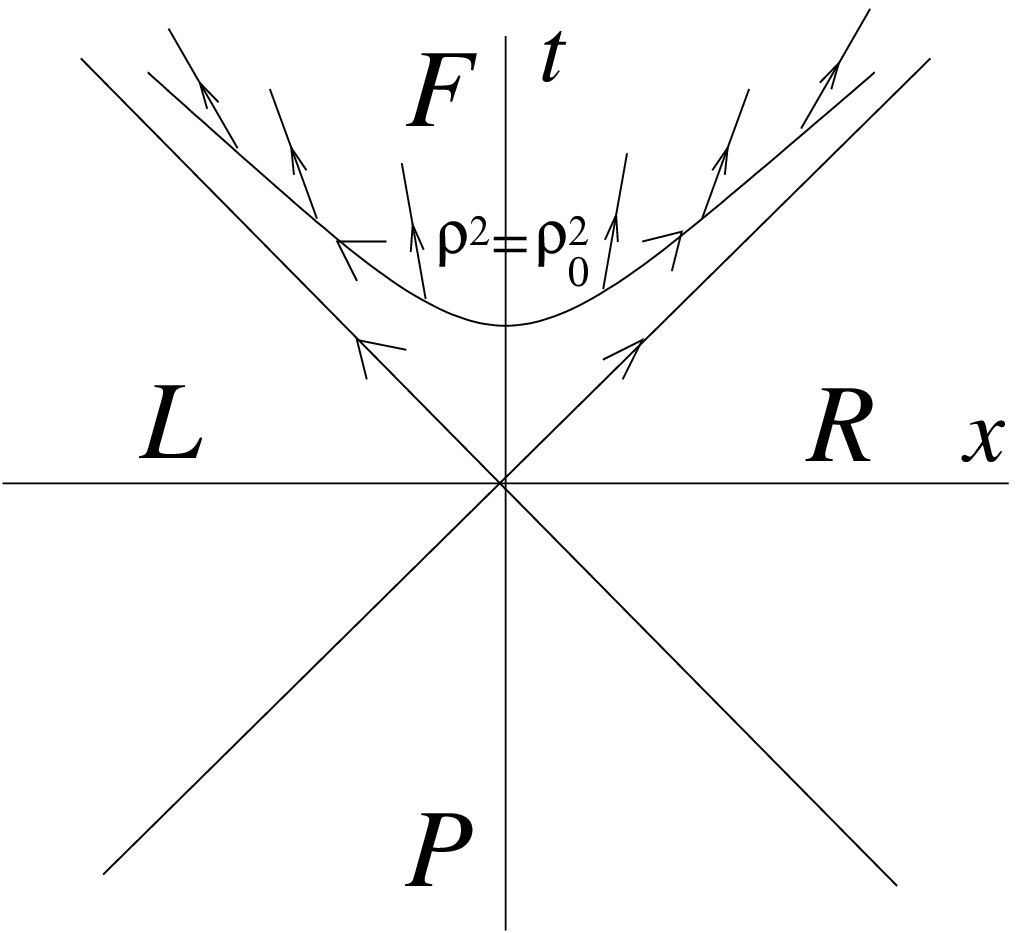}
\end{minipage}\hfill
\begin{minipage}[b]{.46\linewidth}
\includegraphics[width=8.3cm]{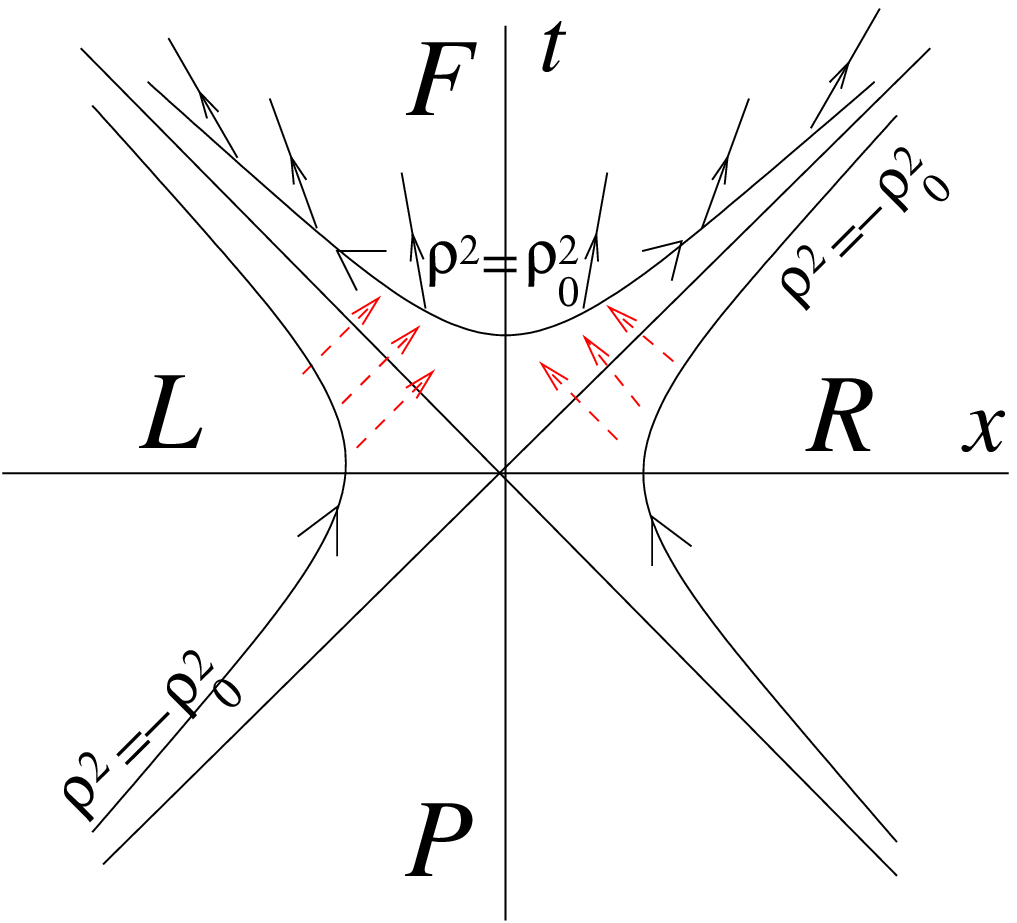}
\end{minipage}
\caption{
The space--time picture of a relativistic heavy ion collision. Left: heavy ions approach the interaction region 
around $x = t = 0$ along the light cone from $x = t = - \infty $ and $x = -t = \infty$. The collision at $x = t = 0$ 
produces the final state particles which approach the mass shell at some proper time $\tau = (t^2 - x^2)^{1/2}$ 
(or, equivalently, along the surface of Rindler space $\rho^2 = \rho^2_0 = x^2 - t^2 < 0$).  The produced particles are 
distributed in rapidity, or in the Rindler coordinate $\eta = {1 \over 2} \ln \left|{t + x \over t - x}\right|$.
Right: also shown in the left (L) and right (R) sectors are the trajectories of space-like $p^2 \sim - Q_s^2 < 0$ 
partons confined in the initial nuclear wave functions characterized by the saturation scale $Q_s$. 
Quantum tunneling of partons from left (L)  and right (R) sectors through this event horizon into the future (F) indicated 
by dashed arrows 
creates a thermal state of parton matter with the temperature $T = Q_s / 2 \pi$.     
}
\label{rindler}
\end{figure}

The conventional space--time picture of a relativistic heavy ion collision is depicted in the left panel of Fig.\ref{rindler}. 
The colliding heavy ions approach the interaction region along the light cone from $x = t = - \infty $ and $x = -t = \infty$. 
The partons inside the nuclei in the spirit of the collinear factorization approach are assumed to have a vanishing 
transverse momentum $k_T$, have a zero virtuality $k^2 = - k_T^2 =0$, and thus are also localized on the light cone 
at $ \pm x = t$. The collision at $x = t = 0$ produces the final state particles with transverse momenta $p_T$ which according 
to the uncertainty principle approach their mass shell at a proper time $\tau = (t^2 - x^2)^{1/2} = \tau_0 \sim 1/p_T$. 
\vskip-1cm
%\vspace{1cm}

For further discussion it is convenient to introduce the Rindler coordinates
\beq\label{rcoord}
\rho^2 = x^2 - t^2; \hspace{1cm} \eta = {1 \over 2} \ln \left|{t + x \over t - x } \right|;
\eeq
The surface of a fixed proper time which is a hyperbola in Minkowski space thus represents a line at $\rho^2 = x^2 - t^2 = \rho^2_0 < 0$ in Rindler space. The Rindler coordinate $\eta$ in high energy physics is often called a space--time rapidity. 

Consider now the case when partons in the wave functions of the colliding nuclei have non-vanishing transverse momenta, as in the Color Glass 
Condensate picture where their transverse momenta are on the order of the saturation scale $Q_s$. In this case 
the partons are space--like $k^2 = - k_T^2$ and are located off the light cone. 
As the colliding nuclei approach each other, these partons begin to interact; note that since they are space--like, 
their interactions are acausal, and are responsible for the breakdown of factorization for the parton modes with $k_T \leq Q_s$.  The interactions of partons with the color fields $g E \simeq Q_s^2$ of another nucleus decelerate them, with a typical acceleration $|a| \simeq Q_s$. The space--time trajectories of the partons are thus given by hyperbolae in Minkowski space, or by the lines with a fixed value of $\rho^2 = - \rho^2_0 = Q_s^2 > 0$ in Rindler space. These trajectories are shown on the right panel of Fig.\ref{rindler}. For partons moving in the left (L) 
and right (R) sectors of space--time with an 
acceleration $|a| = \rho^{-1}$ the light cone surfaces $\rho^2 = 0$, $\eta = \pm \infty$ represent the event horizon 
of the future (F). The information from the future is hidden from them, and the sector F is classically disconnected 
from L and R. However, as discussed above, the future sector F can be reached from 
the left L and right R sectors by quantum tunneling.

\subsection{Hawking--Unruh phenomenon in the parton language}

It is important to mention that Fig.\ref{rindler} does not imply any "bouncing" of the colliding nuclei -- the on--shell particles, as 
well as the high--momentum valence partons from the colliding nuclei are transparent for each other. The soft components 
of the parton wave functions however do interact strongly.  Nevertheless, the whole picture at first glance looks completely 
orthogonal to the conventional parton model. Indeed, in parton model the Weizs{\"a}cker--Williams gluon fields surrounding 
the valence quarks are transverse, with $\vec{E} \perp \vec{H} \perp \vec{p}$ ($\vec{p} \ $ is the momentum of the quark). 
The corresponding "equivalent gluons" are almost on mass shell, and no longitudinal chromo--electric field is present: 
the gluon field tensor is flat in the longitudinal direction: $F_{+-}=0$ ("$+$" and "$-$" refer to the light--cone components).

However, a more careful analysis reveals that this picture is not complete: the configuration of the gluon field produced 
in the sector "F" of  Fig.\ref{rindler} is characterized by $F_{+-} \neq 0$, with a substantial longitudinal chromo--electric field $E_z$. 
In the Color Glass Condensate picture, the strength of the field is  $E_z \sim Q_s^2/g$, and the duration of the pulse is $\sim Q_s^{-1}$. 
Note that in the conventional string model picture, the produced chromo--electric field is purely longitudinal, with the strength 
proportional to the string tension. This exhibits a possible continuity between the string and parton approaches to multi--particle 
production, and suggests the existence of the minimal allowed value of the saturation momentum $Q_{s\ min}$. Basing on the arguments given above and \eq{hag}, one is led to the conclusion that 
\beq\label{qsmin}
Q_{s\ min} = 2\pi T_{Hagedorn} \simeq 1\ {\rm GeV}.
\eeq
A quantitative analysis of parton production in the longitudinal chromo--electric field performed in the framework of the Color 
Glass Condensate approach is underway \cite{KLT}.

\section{SUMMARY} 

In this talk I have argued that the statistical features of multi--particle production may emerge as a consequence of 
the Hawking--Unruh effect. The acceleration, and the emergence of the corresponding event horizon for partons, is caused by 
the pulse of chromo--electric field which accompanies inelastic interactions at high energies.  For hadron interactions at moderate energies, the 
effective temperature appears equal to the Hagedorn temperature \eq{hag}. Once the strength of the chromo--electric field $E \sim Q_s^2/g$ exceeds a critical value determined by \eq{qsmin}, the partons are produced with an effective temperature $T>T_{Hagedorn}$, i.e. in a deconfined state. The subsequent evolution of the produced partonic system has to be taken into account in this case. 

\vskip0.3cm

I am grateful to K. Tuchin for sharing the fun of thinking about the problem discussed here, and to T. Cs{\"o}rgo, G. Dunne, R. Glauber, E. Levin, L. McLerran, G. Nayak  
and R. Venugopalan for helpful discussions.

\end{document}